\documentclass[11pt]{article}
\usepackage{times}
\usepackage{sidecap}
\usepackage{geometry}
\usepackage{floatrow}
\usepackage{float}
\usepackage{hyperref}
\usepackage{graphics,graphicx}
\usepackage{subcaption}
\usepackage{epsf,wrapfig}
\usepackage{stfloats}
\usepackage{marvosym}
\usepackage[dvipsnames]{xcolor}

\usepackage{natbib,bm}
\usepackage{caption}
\usepackage{pdflscape}

\usepackage{paralist}
\let\olditem\item
\renewenvironment{thebibliography}[1]{%
  \section*{\refname}
  \let\par\relax
  \renewcommand{\item}[1][]{\olditem[\textbullet]}%
  \inparaenum}{\endinparaenum}

\newcounter{aff}

\DeclareGraphicsExtensions{.jpg,.pdf,.png,.eps}

\usepackage{caption}
\usepackage[dvipsnames]{xcolor}

\usepackage{tcolorbox}

\bibpunct{(}{)}{;}{a}{}{,}

\usepackage{soul}
\usepackage{float}

\definecolor{darkcyan}{rgb}{0.0, 0.0, 0.7}

\definecolor{boxcolor}{RGB}{193, 229, 255}
\definecolor{figboxcolor}{RGB}{193, 229, 255}
\definecolor{boxcolorgray}{RGB}{227, 227, 227}
\definecolor{darkcyan}{rgb}{0.0, 0.0, 0.7}

\sethlcolor{boxcolorgray}

\def\gtrsim{\lower 2pt \hbox{$\, \buildrel {\scriptstyle >}\over
{\scriptstyle \sim}\,$}}
\def\lesssim{\lower 2pt \hbox{$\, \buildrel {\scriptstyle <}\over
{\scriptstyle \sim}\,$}}

\def\approxlt{\lower.2em\hbox{$\buildrel < \over \sim$}}
\def\approxgt{\lower.2em\hbox{$\buildrel > \over \sim$}}

\def\be{\begin{equation}}
\def\ee{\end{equation}}
\def\bea{\begin{eqnarray*}}
\def\eea{\end{eqnarray*}}

\def\jref@jnl#1{{\rm#1}}
\def\aj{\jref@jnl{AJ}}			
\def\araa{\jref@jnl{ARA\&A}}		
\def\apj{\jref@jnl{ApJ}}			
\def\apjl{\jref@jnl{ApJ}}		
\def\apjs{\jref@jnl{ApJS}}		
\def\ao{\jref@jnl{Appl.~Opt.}}		
\def\apss{\jref@jnl{Ap\&SS}}		
\def\aap{\jref@jnl{A\&A}}		
\def\aapr{\jref@jnl{A\&A~Rev.}}		
\def\aaps{\jref@jnl{A\&AS}}		
\def\azh{\jref@jnl{AZh}}			
\def\baas{\jref@jnl{BAAS}}	
\def\icarus{\jref@jnl{Icarus}}
\def\jrasc{\jref@jnl{JRASC}}		
\def\memras{\jref@jnl{MmRAS}}		
\def\mnras{\jref@jnl{MNRAS}}		
\def\pra{\jref@jnl{Phys.~Rev.~A}}	
\def\prb{\jref@jnl{Phys.~Rev.~B}}	
\def\prc{\jref@jnl{Phys.~Rev.~C}}	
\def\prd{\jref@jnl{Phys.~Rev.~D}}	
\def\pre{\jref@jnl{Phys.~Rev.~E}}	
\def\prl{\jref@jnl{Phys.~Rev.~Lett.}}	
\def\pasp{\jref@jnl{PASP}}		
\def\pasj{\jref@jnl{PASJ}}		
\def\qjras{\jref@jnl{QJRAS}}		
\def\skytel{\jref@jnl{S\&T}}		
\def\solphys{\jref@jnl{Sol.~Phys.}}	
\def\sovast{\jref@jnl{Soviet~Ast.}}	
\def\ssr{\jref@jnl{Space~Sci.~Rev.}}	
\def\zap{\jref@jnl{ZAp}}			
\def\nat{\jref@jnl{Nature}}		
\def\iaucirc{\jref@jnl{IAU~Circ.}}
\def\aplett{\jref@jnl{Astrophys.~Lett.}}
\def\apspr{\jref@jnl{Astrophys.~Space~Phys.~Res.}}
\def\bain{\jref@jnl{Bull.~Astron.~Inst.~Netherlands}}
\def\fcp{\jref@jnl{Fund.~Cosmic~Phys.}}
\def\gca{\jref@jnl{Geochim.~Cosmochim.~Acta}}
\def\grl{\jref@jnl{Geophys.~Res.~Lett.}}
\def\jcp{\jref@jnl{J.~Chem.~Phys.}}	
\def\jgr{\jref@jnl{J.~Geophys.~Res.}}	
\def\jqsrt{\jref@jnl{J.~Quant.~Spec.~Radiat.~Transf.}}
\def\memsai{\jref@jnl{Mem.~Soc.~Astron.~Italiana}}
\def\nphysa{\jref@jnl{Nucl.~Phys.~A}}
\def\physrep{\jref@jnl{Phys.~Rep.}}
\def\physscr{\jref@jnl{Phys.~Scr}}
\def\planss{\jref@jnl{Planet.~Space~Sci.}}
\def\procspie{\jref@jnl{Proc.~SPIE}}

\def\ion#1#2{#1$\;${\small\rm\@Roman{#2}}\relax}

\newcount\lecurrentfam
\def\LaTeX{\lecurrentfam=\the\fam \leavevmode L\raise.42ex
\hbox{$\fam\lecurrentfam\scriptstyle\kern-.3em A$}\kern-.15em\TeX}
\def\sizrpt{
(\fontname\the\font): em=\the\fontdimen6\font, ex=\the\fontdimen5\font
\typeout{
(\fontname\the\font): em=\the\fontdimen6\font, ex=\the\fontdimen5\font
}}
\definecolor{dark}{rgb}{ 0.34, 0.09, 0.49}

\usepackage{fancyhdr,url}
\usepackage{transparent}
\usepackage{tablefootnote}
\usepackage{mathpazo}
\usepackage{tabularx}
\usepackage{booktabs}
\usepackage{longtable}
\usepackage{booktabs}
\usepackage{tabularx}
\usepackage{threeparttable}
\usepackage{enumitem}

\begin{document}

\noindent
\begin{raggedright}
{\Huge\bfseries From local chemistry to galactic evolution: unlocking the power of interstellar isotopes with ALMA2040\\[0.5em]
}\end{raggedright}
\noindent
\vspace{-0.5cm}

\textbf{Laura Colzi}$^1$\textsuperscript{\Letter},
Hayley A. Bunn$^2$,
Víctor M. Rivilla$^1$,
Beatrice Kulterer$^3$,
Eleonora Bianchi$^4$,
Maryvonne Gerin$^5$,
Pamela Klaassen$^6$,
Maria Teresa Beltrán$^4$,
Jes K. Jørgensen$^7$,
Claudio Codella$^4$,
Martin Cordiner$^{8,9}$,
Audrey Coutens$^{10}$, 
Timea Csengeri$^{11}$,
Marta De Simone$^{12,4}$,
Maria N. Drozdovskaya$^{13}$,
Ana Duarte-Cabral$^{14}$,
Francesco Fontani$^4$,
Pablo García$^{15,16}$,
Xing Lu$^{17}$,
Sébastien Maret$^{18}$,
Sergio Martín$^{19,20}$,
Kazi L. J. Rygl$^{21}$, 
Giovanni Sabatini$^4$, 
Nami Sakai$^{22}$,
Álvaro Sánchez-Monge$^{23}$,
Miguel Sanz-Novo$^2$, 
Claudia Toci$^{24}$,
Shaoshan Zeng$^{22}$,
Aurore Bacmann$^{19}$,
Maria J. Jimenez-Donaire$^{25,26}$,
Miriam G. Santa-Maria$^{27}$,
Alessio Traficante$^{28}$

\vspace{1em}
\noindent {\footnotesize{\textit{(Affiliations can be found after the references)}}}

\vspace{+0.3cm}
\begin{tcolorbox}[boxsep=3pt,left=0pt,right=0pt,top=1pt,bottom=1pt,colframe=boxcolor,colback=boxcolor]
\justify
\textbf{Abstract:}
Molecular isotopic ratios provide a unique window into the chemical evolution of the interstellar medium, linking stellar nucleosynthesis and Galactic chemical evolution to the physical and chemical processes shaping molecular clouds, star-forming regions, and planetary systems. However, molecular isotopic ratios do not necessarily trace the underlying elemental composition, as chemical isotopic fractionation, isotope-selective photodissociation, grain chemistry, and radiative-transfer effects can significantly alter the observed values. Disentangling these effects requires simultaneous observations of multiple transitions, isotope systems and molecular families, including rare, doubly, and multiply substituted isotopologues involving D, $^{13}$C, $^{15}$N, $^{17,18}$O, $^{29,30}$Si, and $^{33,34,36}$S (e.g., $^{13}$C$^{15}$N, $^{13}$C$^{34}$S, and $^{13}$C$^{18}$O). ALMA2040, with an order-of-magnitude improvement in sensitivity and at least 512~GHz of instantaneous bandwidth, will make these species accessible across statistically significant samples of Galactic and nearby extragalactic sources.

By enabling systematic isotopic inventories from molecular clouds and prestellar cores to protoplanetary systems, across the Milky Way and in external galaxies, ALMA2040 will establish molecular isotopes as quantitative probes of chemical inheritance, stellar nucleosynthesis, and galaxy evolution, disentangling local chemical processing from the large-scale evolution of matter from stars to new planetary systems.
\end{tcolorbox}

\newpage{}

\begin{figure}[H]
    \centering
    \includegraphics[width=0.9\textwidth]{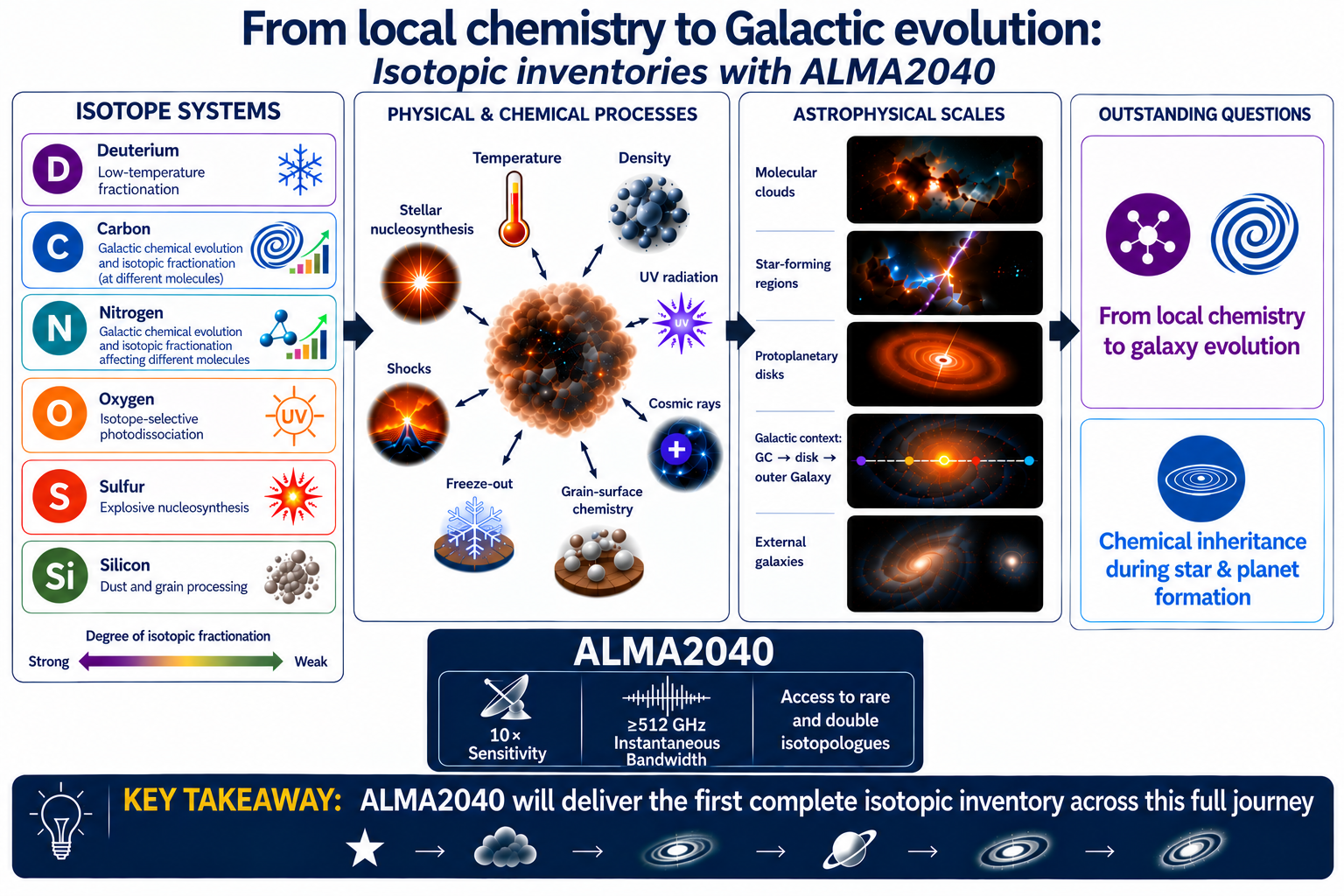}
    \caption{\small Schematic overview of isotopic ratios in the ISM and their connection to chemical fractionation and galactic chemical evolution. Symbols indicate the main drivers for each isotope system: low-temperature fractionation (D), GCE, and isotopic fractionation for different molecular tracers (C and N), isotope-selective photodissociation (O), explosive nucleosynthesis (S), and dust and grain processing (Si). The figure highlights the new science enabled by ALMA2040.}
    \label{fig:isotopes_overview}
\end{figure}

\vspace{-0.7cm}
\section{Scientific context and open questions}

\vspace{-0.2cm}
Isotopic ratios record the nucleosynthesis and chemical history of the interstellar medium (ISM), from au-scale planet-forming regions to kpc-scale galaxies and from cloud evolution ($\sim10^{4-5}$~yr) to Galactic chemical evolution ($\sim10^9$~yr).

An isotopologue is a molecule containing one or more isotopes of its constituent elements. Isotopologues containing a less abundant isotope are referred to as rare isotopologues; when two or more atoms in a molecule are replaced by rare isotopes, we refer to them as doubly substituted or multiply substituted isotopologues, respectively.
Different isotope systems, i.e. groups of isotopes of individual elements, trace a diversity of physical and chemical processes:\\ \vspace{-0.0cm}\textbf{$\bullet$ Deuterium (D/H):} Strongly enhanced in molecules within cold ($T\lesssim20$ K), dense ($n_{\rm H_{2}}\sim$ 10$^{5}$ cm$^{-3}$) molecular clouds through low-temperature isotope-exchange reactions, making deuterated molecules sensitive to temperature, ionization, and CO depletion during early star formation (\citealt{caselliceccarelli2012,colzi2022a,sabatini2024}).\\
\textbf{$\bullet$ Carbon ($^{12}$C/$^{13}$C):} Increases from $\sim40$ in the Galactic Center to $\gtrsim100$ in the outer Galaxy, tracing stellar nucleosynthesis and galactic chemical evolution (GCE) while also being affected by molecular isotope-selective chemistry (\citealt{milam2005,visser2009,sakai2013,romano2017,colzi2020,colzi2026}).\\
\textbf{$\bullet$ Nitrogen ($^{14}$N/$^{15}$N):} Varies from $\sim100$ to $>1000$ across the ISM and Solar System, reflecting the combined effects of nucleosynthesis, isotope-selective chemistry, and inheritance during the star formation process (\citealt{heays2014,hily-blant2017,furuya2018,colzi18a,colzi18b,colzi2019,colzi2022b}).\\
\textbf{$\bullet$ Oxygen ($^{16}$O/$^{17}$O/$^{18}$O):} Probes massive-star nucleosynthesis and Galactic enrichment, while isotope-selective photodissociation and CO self-shielding make O-isotopic ratios sensitive to UV irradiation and cloud structure (\citealt{visser2009,romano2017}).\\
\textbf{$\bullet$ Sulfur ($^{32}$S/$^{33}$S/$^{34}$S/$^{36}$S):} Traces massive-star and supernova nucleosynthesis and is expected to experience relatively weak chemical fractionation, providing a valuable probe of intrinsic Galactic isotopic composition (\citealt{kobayashi2011,loison2019b,yan2023}).\\
\textbf{$\bullet$ Silicon ($^{28}$Si/$^{29}$Si/$^{30}$Si):} Constrains nucleosynthesis and GCE, while Si isotopologues additionally trace dust processing and shock-driven grain destruction (\citealt{draine1980,schilke1997,kobayashi2020}).


Isotope systems respond differently to local chemistry: silicon and sulfur are weakly fractionated, deuterium can be enriched by orders of magnitude, and C, N, and O show intermediate behavior. \textbf{Measuring several isotope systems in the same sources is essential to separate local chemistry from Galactic evolution.} Isotopic substitution can also induce weak dipole moments in otherwise radio-inactive apolar molecules, such as H$_3^+$ and C$_2$H$_2$, making some isotopologues accessible through rotational spectroscopy (Table~\ref{tab:isotopologues}).
 
Broadband IRAM 30m and Yebes 40m surveys have detected faint multiply substituted isotopologues, including H$^{13}$C$^{13}$CCN, H$^{13}$C$^{15}$N, and D$^{13}$CCCN \citep{gerin2006,gerin2009,belloche2013,tercero2024,colzi2026}. Their rarity limits statistical tests of chemical versus environmental variations. ALMA2040's $\sim10\times$ line-sensitivity gain and broader bandwidth will extend such measurements to representative samples of Galactic and extragalactic sources.

\textbf{Outstanding open questions:}\\
\textbf{\color{RedViolet} (1) From local chemistry to galaxy evolution:}
How can stellar nucleosynthesis, Galactic chemical evolution (GCE), local isotopic fractionation, and radiative-transfer effects be disentangled in observed isotopic ratios? Which molecular species best trace elemental isotopic abundances, and how do multi-isotope ratios (H, C, N, O, Si, and S) vary across the Milky Way and external galaxies? Can these ratios serve as quantitative tracers of nucleosynthesis, gas mixing, metallicity, star-formation history, and nuclear activity?\\
\textbf{\color{RedViolet} (2) Chemical inheritance during star and planet formation:}
What processes regulate isotopic ratios during star and planet formation, particularly the strong deuterium enrichment in cold clouds? To what extent are isotopic signatures in disks, comets, and primitive Solar System bodies inherited from their parent clouds or modified during protoplanetary disk evolution?

Figure~\ref{fig:isotopes_overview} summarizes these connections, while Table~\ref{tab:isotopologues} (after the references) compiles rare, doubly, and multiple isotopologues with available rotational spectroscopy, highlighting the molecular inventory accessible to ALMA2040. This inventory is expected to expand substantially over the next 15 years as high-resolution rotational spectroscopy techniques continue to advance. The following science cases demonstrate how these tracers can address the fundamental questions above.

\vspace{-0.6cm}
\section{Science Cases for ALMA2040}
\vspace{-0.2cm}
\subsection{Science Case I: From local isotopic chemistry to the chemical evolution of galaxies}
\vspace{-0.2cm}

\textbf{ \color{WildStrawberry} The Milky Way.} Stellar nucleosynthesis enriches the ISM with stable isotopes, whose ratios (C, N, O, S, and Si) trace the Milky Way’s chemical history \citep{wilson1994,kobayashi2020,romano2022}. However, fractionation, selective photodissociation, and freeze-out also alter molecular isotope ratios locally \citep{roueff2015,colzi2020,sipila2023}. Disentangling Galactic chemical evolution from local chemistry is therefore a major challenge. The elemental isotope ratio set by stellar nucleosynthesis and GCE can be expressed schematically as   
\setlength{\abovedisplayskip}{0pt}
\setlength{\belowdisplayskip}{0pt}
\setlength{\abovedisplayshortskip}{0pt}
\setlength{\belowdisplayshortskip}{0pt}
\begin{equation}
R_{\rm GCE} =
R_{\rm obs}\,
f_{\rm chem}\,
f_{\tau}\,
f_{\rm ex},
\end{equation}
where $R_{\rm obs}$ is the observed molecular isotope ratio, while $f_{\rm chem}$, $f_{\tau}$, and $f_{\rm ex}$ describe modifications due to local chemistry, optical depth, and excitation, respectively \citep{watson1976,langer1984,goldsmith1999,lis2010,colzi2026}.
Current observations generally cannot independently constrain these factors. 
They rely mainly on the main and singly substituted isotopologues, whose emission is often optically thick and affected by uncertain excitation, resulting in systematic uncertainties of $\sim20$--$50\%$, and up to a factor of two in some cases. These uncertainties are comparable to, or larger than, the $\sim10$--$30\%$ Galactic variations expected from stellar nucleosynthesis over much of the Galactic disk (e.g., \citealt{milam2005,adande2012,colzi18b,colzi2022b,colzi2026,yan2023}). ALMA2040 will address this limitation by enabling routine detection of rare, double, and multiple isotopologues, which provide substantially less optically thick constraints on isotopic ratios ($f_{\tau}\rightarrow1$). Importantly, the more abundant isotopologues of a rare pair may remain optically thick, so robust ratios will require observations of multiple isotopologues and transitions to constrain optical-depth and excitation effects. The targeted instantaneous bandwidth of $\geq512$~GHz will enable simultaneous observations of multiple isotopologues and transitions across a broad frequency range, substantially improving constraints on excitation and optical depth and reducing the associated uncertainties ($f_{\rm ex}\rightarrow1$) \citep{tercero2024,colzi2026}. Combined with state-of-the-art astrochemical models and laboratory measurements including multiple isotopologues \citep{loison2019b,colzi2020,sipila2023,jimenez-redondo2025}, this will enable more robust separation of optical-depth, excitation, and chemical-fractionation effects and measurements of the underlying elemental isotope ratios. For sources with sufficiently strong detections, these advances could reduce the observational uncertainties in molecular isotopic ratios from the current $\sim20$--$50\%$ to $\sim10$--$20\%$, an improvement of approximately a factor of 2--3. The accuracy of the inferred elemental isotope ratios will additionally depend on how well local chemical fractionation, $f_{\rm chem}$, can be constrained through observations and modeling.

\begin{figure}[H]
    \centering
    \includegraphics[width=0.9\linewidth]{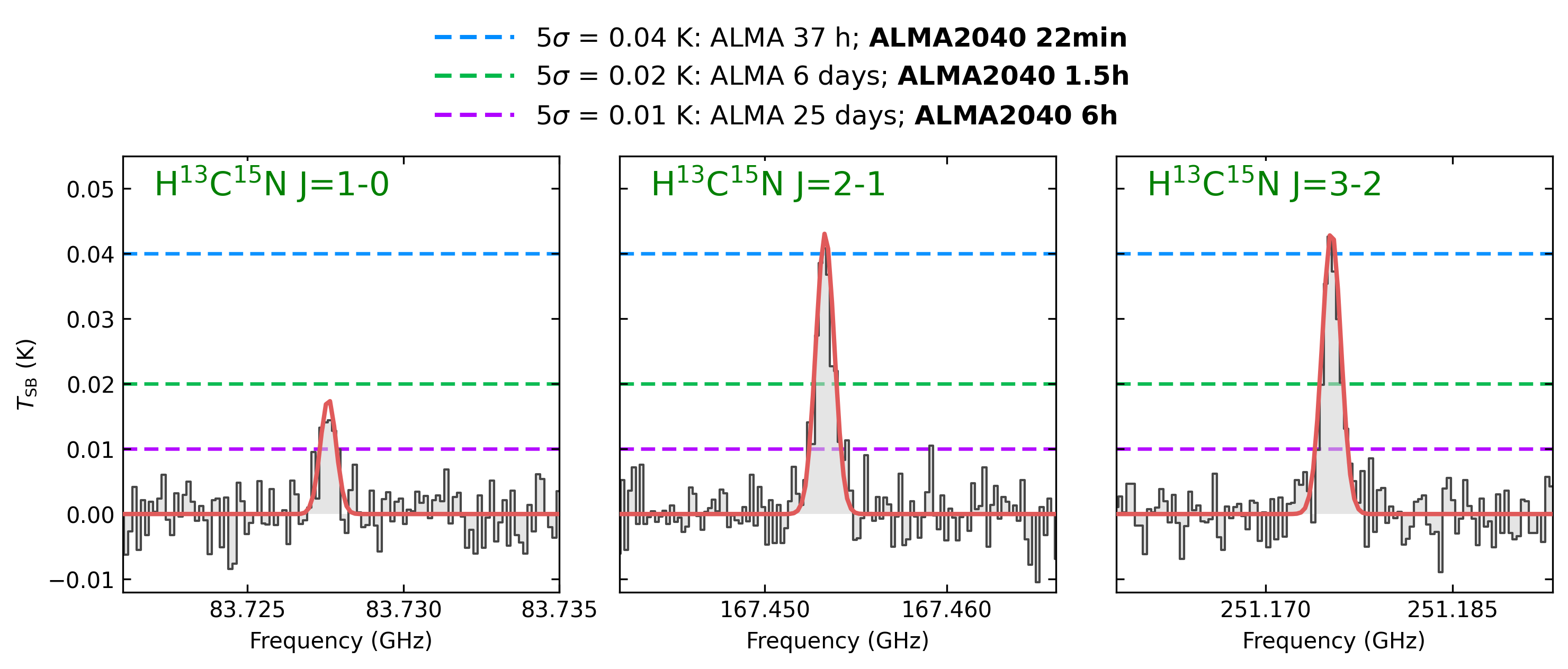}
    \caption{\small Simulated H$^{13}$C$^{15}$N $(1-0)$, $(2-1)$, and $(3-2)$ transitions toward a typical massive star-forming region. The simulations are based on HC$^{15}$N $(1-0)$ observations obtained with NOEMA at $3^{\prime\prime}$ angular resolution (private communication). We adopt a $^{12}$C/$^{13}$C ratio of 83, appropriate for a galactocentric distance of 10 kpc \citep{colzi2026}, $T_{\rm ex}=15$ K, and a FWHM linewidth of 2 km s$^{-1}$. An rms noise of 3.6 mK is required for a $5\sigma$ detection of H$^{13}$C$^{15}$N $(1-0)$. The $\times10$ sensitivity increase of ALMA2040 will enable multiple transitions to be observed in a single source within a reasonable amount of time.}
    \label{fig:simulation1}
\end{figure}

Species such as H$^{13}$C$^{15}$N, $^{13}$C$^{34}$S, $^{13}$C$^{18}$O, HC$^{15}$N, H$^{13}$CN, $^{29}$SiO, and $^{30}$SiO (see Fig.~\ref{fig:simulation1} and Table~\ref{tab:isotopologues}) will provide complementary constraints on elemental isotope ratios. Combining multiple isotopologues and transitions will allow optical-depth and excitation effects to be quantified rather than assuming that the relevant lines are optically thin. Simultaneous observations of multiple molecular families (HCN, HNC, CN, CO, CS, SO, SiO, OCS, etc.) will distinguish molecule-dependent fractionation from genuine Galactic abundance variations.
In the Milky Way, ALMA2040 will enable systematic mapping of massive star-forming regions from diffuse envelopes to dense cores, resolving $\sim0.05$~pc ($\sim10^4$~au) scales within $\sim0.5$~pc structures, thereby linking isotopic variations across the cloud scales relevant to star formation. Isotope ratios can thus be measured as a function of density, temperature, UV irradiation, and evolutionary stage within the same molecular complex, directly quantifying local chemical effects \citep{colzi2019}. \textit{Large surveys of hundreds of sources from the Galactic Center through the spiral arms to the outer Galaxy will provide a statistical view of the Milky Way's elemental isotopic composition, constraining stellar nucleosynthesis, radial gas mixing, and GCE }(\citealt{kobayashi2020,colzi2022b}, see also Fontani et al. ALMA2040 white paper).

\textbf{ \color{WildStrawberry} External galaxies.} The same methodology can be extended to external galaxies. Their isotopic composition encodes their star formation histories, stellar populations, gas accretion, and chemical enrichment, yet extragalactic isotope studies remain largely limited to CO isotopologues and a few exceptionally bright species \citep{henkel1993,aalto1991b,aladro2013,wallstrom2016,martin2021}. The ALMA large program ALCHEMI demonstrated the potential of unbiased molecular surveys to reveal the chemical complexity of galactic nuclei, but isotopic studies remain restricted to the brightest molecules or to single sources (\citealt{martin2019b,martin2021} and references therein, \citealt{colzi2026}).
The order-of-magnitude sensitivity increase and unprecedented instantaneous bandwidth of ALMA2040 will enable the transition from simple molecular inventories to detailed isotopic studies of external galaxies, including high-$z$ systems. ALMA2040 will enable spatially resolved measurements of multiple
isotopic ratios, including $^{12}$C/$^{13}$C,
$^{14}$N/$^{15}$N, $^{16}$O/$^{18}$O, and
$^{32}$S/$^{34}$S, probing molecular gas on scales of
$\sim$50--100 pc in galaxies at distances of 20--50 Mpc
and potentially reaching parsec scales in the nearest galaxies. Multiple isotopologues (currently limited to isolated detections, \citealt{martin2019}) will become accessible across statistically significant samples of nearby spirals, starbursts, AGN, dwarf galaxies, and (U)LIRGs. Multi-transition observations and chemical modeling will simultaneously constrain excitation and local fractionation, while observations of multiple molecular families will probe different nucleosynthetic channels.
Combined with GCE models, these measurements will constrain star formation histories, gas accretion and enrichment timescales, and the relative contributions of different stellar populations. In particular, CNO isotopes provide sensitive constraints on the stellar initial mass function (IMF), since they are produced by stars spanning different masses and evolutionary timescales. Multi-isotope measurements can therefore test predictions of top-heavy IMFs in starburst environments while reducing degeneracies with star formation history and local chemistry \citep{romano2017,zhang2018}.

\vspace{-0.4cm}
\subsection{Science Case II: Tracing the inheritance of isotopic signatures from molecular clouds to planetary systems}
\vspace{-0.2cm}

Molecular inventories reveal which species are present, whereas isotopic compositions provide complementary information on where, when, and through which pathways they formed. A major open question is whether these isotopic signatures are inherited from parent molecular clouds or substantially reprocessed during star and planet formation. Deuterium provides the clearest example, with molecular D/H ratios enhanced by more than four orders of magnitude above the elemental abundance ($\sim2.5\times10^{-5}$) in cold ($T\lesssim20$ K), dense gas. Similar C and N isotopic anomalies are observed from prestellar cores to protostars, disks, comets, and meteorites, suggesting that part of a planetary system's isotopic inventory may survive from the earliest stages of star formation. Observations of multiple isotopologues across evolutionary stages are therefore essential for distinguishing inheritance from local reprocessing.

\textbf{ \color{WildStrawberry} Deuteration as a probe of chemical inheritance.} Deuteration records the physical and chemical history of star-forming material. For grain-surface products such as methanol, the H$_2$D$^+$/H$_3^+$ ratio regulates H/D incorporation into ices and depends strongly on temperature and chemical conditions \citep{tielens1983}. Different deuteration levels can therefore trace successive stages of ice formation and potentially act as chemical clocks \citep{fontani2011}. In gas-phase molecules, deuteration is instead governed by kinetic isotope effects, providing complementary constraints on formation pathways that can be tested with chemical models \citep{roueff2007,skouteris2017,riedel2025}.
Methanol is currently the best benchmark. Several isotopologues, including CH$_2$DOH, CH$_3$OD, CHD$_2$OH, and CD$_3$OH, have been detected, with CD$_3$OD recently detected toward the protostar IRAS~4A (e.g., \citealt{jacq1993detection,parise2002detection,parise2004first,Lin23,ilyushin2023investigation,belloche2026}). Non-statistical D incorporation demonstrates that isotopic patterns can retain information on molecular formation pathways \citep{jorgensen2018}, while laboratory spectroscopy continues to expand and improve the frequency and intensity data required for deuterated isotopologues and complex organic molecules (COMs), particularly for the highly excited states populated in warm gas \citep{su1991microwave,margules2023rotational,oyama2023,chen2026isotopic}. Extending such studies to S- and N-bearing species, such as CH$_{3}$SH, whose D-isotopologues are extremely faint, will provide complementary constraints on chemical inheritance \citep{zakharenko2019deuterated,bunn2025laboratory}.

Prestellar cores provide the initial conditions for this fractionation. Their low temperatures ($\lesssim20$ K) and high densities ($\gtrsim10^4$~cm$^{-3}$) favour strong deuteration, and methanol is currently the only COM for which several deuterated isotopologues have been securely detected in prestellar cores, with D/H ratios comparable to those measured in low-mass protostars \citep{Scibelli25,Kulterer26}. This supports prestellar inheritance and is consistent with cometary methanol measurements and chemical models favouring prestellar formation followed by inheritance \citep{Drozdovskaya21,Kulterer22}. Other COMs, including acetaldehyde, methyl formate, and methyl cyanide, are detected in prestellar environments, but their deuterated isotopologues remain below current sensitivity limits \citep{Scibelli20,Scibelli24}. While prestellar cores can be studied with sensitive single-dish observations, ALMA2040 will combine substantially improved sensitivity with high angular resolution, enabling more efficient detection of faint isotopologues and their spatially resolved characterization from prestellar cores through protostars and disks. In particular, detecting multiply deuterated COMs beyond the handful of exceptionally bright Class~0 sources currently accessible, such as IRAS~4A and IRAS~16293--2422, will test whether the high deuteration observed in these benchmark objects is representative of low-mass star formation. Spatially resolving rare isotopologues will reveal how inherited chemical signatures evolve from the envelope to the emerging disk.

\textbf{\color{WildStrawberry} Beyond deuterium.} $^{13}$C and $^{15}$N provide complementary tracers of molecular inheritance and local processing, with evidence that isotopic compositions may vary between molecular species and chemical reservoirs. Toward IRAS~16293--2422, tentative differences in the $^{12}$C/$^{13}$C ratios of several COMs suggest that carbon isotopic fractionation may not be described by a single elemental ratio \citep{jorgensen2018}. These results are particularly interesting in light of recent JWST measurements of $^{12}$C/$^{13}$C in CO and CO$_2$ ices toward protostellar envelopes, which reveal variations in the solid-state carbon isotopic composition \citep{brunken2024}. Isotopic measurements also require care because the main isotopologue of abundant species such as CH$_3$OH can be highly optically thick, leading to underestimated column densities and biased isotopic ratios. Multiple rare isotopologues are therefore essential to distinguish optical-depth effects from genuine isotopic variations. ALMA2040's sensitivity and broad bandwidth will enable such measurements across multiple molecular families and, where possible, different positions within the same molecule.
These signatures may be inherited from the molecular cloud into the disk, but can also be modified by chemical reactions, selective photodissociation, and isotope-dependent transport. HCN observations in protoplanetary disks, for example, find $^{14}$N/$^{15}$N ratios comparable to those measured in CN and NH$_2$ in Solar System comets ($\sim$100--150), with spatial variations within individual disks \citep{hily-blant2013a,guzman2017}. The interstellar object 3I/ATLAS further illustrates the potential of isotopologues to probe Galactic and formation histories: JWST measurements found high water D/H ratio of $(0.95\pm0.06)\%$, while VLT observations of CN yielded $^{12}$C/$^{13}$C$\sim150$ and $^{14}$N/$^{15}$N$\sim360$ \citep{cordiner2026,opitom2026}. Whereas comprehensive isotopic studies have so far been performed for only one comet (67P/Churyumov–Gerasimenko, through Rosetta), ALMA2040 will enable systematic measurements across statistically significant samples, constraining the processing and inheritance of material during planets formation.

\vspace{-0.4cm}
\subsection{The ALMA2040 opportunity}
\vspace{-0.2cm}

ALMA2040's order-of-magnitude sensitivity gain and broad bandwidth will enable spatially resolved surveys of multiple rare isotopologues and transitions. Combined with astrochemical and GCE models, these observations will distinguish local fractionation from nucleosynthesis across the Milky Way and external galaxies, and test whether isotopic signatures survive from molecular clouds into planetary systems.
 

\vspace{-0.6cm}
\section{ALMA2040: Technical Requirements}
\vspace{-0.3cm}

The two science cases require major improvements in sensitivity, bandwidth, frequency coverage, and imaging to observe multiple isotope systems and molecular families across diverse physical and evolutionary environments.

\textbf{ \color{Maroon} $\triangleright$  \color{Maroon} Sensitivity and observing efficiency.}
The main limitation of current ALMA capabilities is the sensitivity required to detect rare isotopologues. For a typical massive star-forming region at a Galactocentric distance of 10~kpc, the expected peak brightness of H$^{13}$C$^{15}$N(1--0) is only $\sim20$~mK (private communication). Detecting this line at S/N$\sim5$ with current ALMA requires more than 7 days for a single transition, making systematic surveys impractical. \textbf{A 10$\times$ sensitivity improvement would reduce this to $\sim2$ hr per source}, enabling multi-transition surveys of hundreds of sources, including the fainter deuterated and multiply substituted species (Fig.~\ref{fig:simulation1}). 
As an illustrative observing programme, a survey of 100 Galactic star-forming regions would require $\sim200$~hr in total for the H$^{13}$C$^{15}$N(1--0) benchmark. The actual observing time would depend on the faintest targeted lines, source properties, and number of frequency setups (see below).
Similar gains are essential for fainter deuterated and multiply substituted species targeted in Science Case II. The increased sensitivity will also make line confusion more important in chemically rich regions (see Rivilla et al. ALMA2040 white paper). \textbf{Broad bandwidth and multi-transition observations will improve rare-isotopologue identification and column-density constraints, while combining low- and high-frequency transitions will provide complementary constraints on optical depth, excitation, and spatial structure.} Improved sensitivity will also enable isotopic studies of rarer species, including F- and Al-bearing molecules, in environments such as PDRs.

\textbf{\color{Maroon}$\triangleright$ Broadband instantaneous frequency coverage.}
Key transitions for constraining isotopic ratios, excitation, and molecular abundances span several ALMA bands, requiring broad frequency coverage (Table~\ref{tab:lines}). The Wideband Sensitivity Upgrade (WSU) will increase instantaneous bandwidth from $\sim16$ to $\sim64$~GHz \citep{carpenter2023}, substantially reducing the number of frequency setups. \textbf{Building on the WSU, the current ALMA2040 concept targets at least 512~GHz of instantaneous bandwidth}, enabling simultaneous observations of multiple isotopologues and transitions across a much broader frequency range. Fully realizing its potential will also require close collaboration with laboratory spectroscopists to provide accurate frequencies and intensities for high-excitation transitions and partition functions including vibrational and torsional states, thereby enabling reliable isotope ratios to be derived in warm gas. This capability is essential for efficiently targeting multiple isotope systems (H, C, N, O, Si, and S) and molecular families, while providing homogeneous constraints on excitation, column densities, and isotopic ratios. It will also enable statistically significant surveys of $\sim100$--1000 sources across the Galactic Centre and disk, as well as prestellar cores, protostars, protoplanetary disks, and comets. Similar requirements for broad frequency coverage and efficient multi-line observations are identified in the other ALMA2040 white paper (e.g. Rivilla et al., Bouvier et al.) and working groups, highlighting the importance of this capability across multiple science themes.

\begin{wraptable}{r}{0.42\textwidth}
\centering
\setcounter{table}{1}
\caption{\small Representative transitions illustrating the broad frequency coverage required for studies of isotopic ratios.}
\label{tab:lines}
\begin{tabular}{lccc}
\hline
Species & Transition & $\nu$ (GHz) & Band\\
\hline
DCO$^+$    & 1--0 & 72.04  & 2\\
DCN        & 1--0 & 72.41  & 2\\
N$_2$D$^+$ & 1--0 & 77.11  & 2\\
NH$_2$D    & $1_{1,1}$--$1_{0,1}$ & 85.93 & 2/3\\
HC$^{15}$N & 1--0 & 86.05 & 2/3\\
H$^{13}$CN & 1--0 & 86.34 & 2/3\\
DCO$^+$ & 2--1 & 144.08 & 4\\ 
DCN & 2--1 & 144.83 & 4\\
HC$^{15}$N & 2--1 & 172.11 & 5\\
H$^{13}$CN & 2--1 & 172.68 & 5\\
HC$^{15}$N & 3--2 & 258.16 & 6\\
H$^{13}$CN & 3--2 & 259.01 & 6\\
\hline
\end{tabular}
\vspace{-0.2cm}
\end{wraptable}


\textbf{ \color{Maroon}$\triangleright$ Spectral resolution.}
The spectral resolution currently available with ALMA is sufficient for these science cases: $\sim0.5$~km~s$^{-1}$ resolves the $\sim2$--5~km~s$^{-1}$ lines typical of massive star-forming regions, while $\sim0.1$~km~s$^{-1}$ is needed for the narrow emission of prestellar cores and comets and benefits protoplanetary-disk studies. The WSU will provide $\sim0.1$~km~s$^{-1}$ across the full instantaneous bandwidth, making sensitivity rather than spectral resolution the main requirement for ALMA2040.

\textbf{\color{Maroon}$\triangleright$ Angular resolution and spatial scales.}
Current ALMA angular resolution is sufficient, with scale requirements varying between science cases. For massive star-forming regions at 1--4~kpc, $\sim3^{\prime\prime}$ probes $\sim0.015$--0.06~pc scales (Science Case I), while nearby low-mass star-forming regions require $<0.03^{\prime\prime}$ to resolve compact protostellar structures and disks (Science Case II; see also Drozdovskaya, Csengeri et al.; Walsh et al. ALMA2040 white papers). \textbf{Spatial resolution separates physical and chemical components, while missing short spacings can bias isotopic ratios when isotopologues have different distributions. Recovering emission on tens-of-arcsecond scales with matched spatial filtering therefore requires combining the main array with large single dishes or dedicated compact arrays.}


\textit{Together, these capabilities will provide the observational foundation to disentangle stellar nucleosynthesis and GCE from local chemical processing, and to determine which isotopic signatures are inherited into newly forming planetary systems.}

\vspace{-0.6cm}
\section{Synergies with current and future facilities}
\vspace{-0.2cm}

ALMA2040 will benefit from complementary facilities across the spectrum. ELT and JWST will constrain stellar populations, metallicities, and star-formation histories, while ELT exoplanet spectroscopy will connect planetary and interstellar compositions. Gaia and future GaiaNIR astrometry will trace stellar populations and Galactic kinematics \citep{gaia2023chemical,hobbs2021}. The ngVLA and SKAO will provide HI, radio recombination lines, and centimetre molecular transitions, including low-$J$ isotopic tracers in distant galaxies \citep{balser2018,bianchi2026,forbrich2026}. Far-infrared PRIMA observations will constrain cooling lines and hydrides \citep{bradford2025,glenn2025}. AtLAST or a comparable large single dish could map wide Galactic fields, recover extended emission, and identify targets for ALMA2040 follow-up \citep{klaassen2025}. Together, these facilities will link isotopic chemistry from molecular clouds to planetary systems and galaxies.

\enlargethispage{2\baselineskip}
\vspace{-0.7cm}
\section*{Acknowledgements}
\vspace{-0.3cm}
We thank the referees for their careful reading of the white paper and their constructive comments and suggestions, which helped improve its clarity and scientific motivation.

This work was developed as part of the broader ALMA2040 community effort. We acknowledge the Lorentz Center and NOVA for supporting the community workshops and coordination activities that helped initiate and shape the ALMA2040 White Paper process. 

The project that gave rise to these results received the support of a fellowship from the "la Caixa” Foundation (ID 100010434). The fellowship code is LCF/BQ/PR25/12110012. L.~C. acknowledges support from the grant PID2022-136814NB-I00 funded by the Spanish Ministry of Science, Innovation and Universities/State Agency of Research MICIU/AEI/10.13039/501100011033 and by ERDF, UE.
\vspace{+0.2cm}

\bibliographystyle{apj_mod}
{\bibliography{bibliography}}


\begin{landscape}
\begin{table}[p]
\centering
\tiny
\setlength{\tabcolsep}{1.5pt}
\renewcommand{\arraystretch}{0.65}
\setcounter{table}{0}
\captionsetup{font=tiny,skip=2pt}
\caption{Molecules for which millimeter-wave spectroscopy of rare,
doubly, and multiply substituted isotopologues is available. An online version of this Table is available at \url{https://doi.org/10.5281/zenodo.22940404}.}
\label{tab:isotopologues}

\begin{tabularx}{\linewidth}{@{}l@{\hspace{3pt}}X@{}}
\toprule
\textbf{Parent molecule} &
\textbf{Available rare, double, and multi-substituted isotopologues in mm spectroscopic catalogs} \\
\midrule

\multicolumn{2}{@{}l}{\textit{Diatomic, small inorganic, and C-, N-, O-, S-bearing molecules}}\\
\midrule

CO &
$^{13}$CO, C$^{17}$O, C$^{18}$O, $^{13}$C$^{17}$O, $^{13}$C$^{18}$O \\

CN &
$^{13}$CN, C$^{15}$N, $^{13}$C$^{15}$N \\

CS &
$^{13}$CS, C$^{33}$S, C$^{34}$S, C$^{36}$S,
$^{13}$C$^{33}$S, $^{13}$C$^{34}$S, $^{13}$C$^{36}$S \\

SO &
$^{33}$SO, $^{34}$SO, $^{36}$SO, S$^{17}$O, S$^{18}$O,
$^{34}$S$^{17}$O, $^{33}$S$^{18}$O, $^{34}$S$^{18}$O \\

O$_2$ &
$^{16}$O$^{17}$O,
$^{16}$O$^{18}$O,
$^{17}$O$_2$,
$^{17}$O$^{18}$O,
$^{18}$O$_2$
\\ 

SiO &
Si$^{17}$O, Si$^{18}$O, $^{29}$Si$^{17}$O, $^{29}$Si$^{18}$O,
$^{30}$Si$^{17}$O, $^{30}$Si$^{18}$O \\

SiS &
$^{29}$SiS, $^{30}$SiS, Si$^{33}$S, Si$^{34}$S, Si$^{36}$S,
$^{29}$Si$^{33}$S, $^{29}$Si$^{34}$S,
$^{30}$Si$^{33}$S, $^{30}$Si$^{34}$S \\

OCS &
O$^{13}$CS, $^{17}$OCS, $^{18}$OCS, OC$^{33}$S, OC$^{34}$S, OC$^{36}$S,
$^{13}$C$^{18}$OS, $^{13}$C$^{34}$S,
$^{17}$OC$^{34}$S, $^{18}$OC$^{34}$S \\

OH &
OD, $^{18}$OH \\

CH$^+$ &
$^{13}$CH$^+$, CD$^+$ \\

H$_2$O &
HDO, H$_2^{17}$O, H$_2^{18}$O, D$_2$O,
HD$^{17}$O, HD$^{18}$O, D$_2^{17}$O, D$_2^{18}$O \\

H$_2$S &
HDS, H$_2^{33}$S, H$_2^{34}$S, H$_2^{36}$S,
D$_2$S, HD$^{34}$S, D$_2^{34}$S \\

NH$_3$ &
NH$_2$D, $^{15}$NH$_3$, NHD$_2$, $^{15}$NH$_2$D, ND$_3$ \\

H$_3^+$ & \textbf{H$_2$D$^+$}, \textbf{D$_2$H$^+$} \\

CH$_3^+$ &
\textbf{CH$_2$D$^+$} \\

NH$_4^+$ &
\textbf{NH$_3$D$^+$} \\

H$_3$O$^+$ &
H$_2$DO$^+$ \\

HCO$^+$ &
DCO$^+$, H$^{13}$CO$^+$, HC$^{17}$O$^+$, HC$^{18}$O$^+$,
D$^{13}$CO$^+$, H$^{13}$C$^{17}$O$^+$,
H$^{13}$C$^{18}$O$^+$, D$^{13}$C$^{18}$O$^+$ \\

N$_2$H$^+$ &
N$_2$D$^+$, $^{15}$NNH$^+$, N$^{15}$NH$^+$,
$^{15}$NND$^+$, N$^{15}$ND$^+$, $^{15}$N$^{15}$NH$^+$ \\

SO$_2$ &
$^{33}$SO$_2$, $^{34}$SO$_2$, $^{36}$SO$_2$,
S$^{17}$OO, S$^{18}$OO, $^{34}$S$^{18}$O$_2$ \\

C$_2$H &
$^{13}$CCH, C$^{13}$CH, C$_2$D \\

C$_2$H$_2$ &
\textbf{HCCD} \\

HCN &
DCN, HC$^{15}$N, H$^{13}$CN, $^{13}$C$^{15}$N,
D$^{15}$CN, D$^{13}$CN, D$^{13}$C$^{15}$N \\

HNC &
DNC, H$^{15}$NC, HN$^{13}$C, D$^{15}$NC, D$^{13}$NC \\

H$_2$CO &
HDCO, $^{13}$CH$_2$O, H$_2$C$^{17}$O, H$_2$C$^{18}$O,
D$_2$CO, HD$^{13}$CO, D$_2^{13}$CO,
H$_2^{13}$C$^{17}$O, H$_2^{13}$C$^{18}$O \\

H$_2$CCO &
H$_2$CC$^{18}$O, H$_2$C$^{13}$CO, HDCCO, D$_2$CCO \\

H$_2$CS &
HDCS, H$_2^{13}$CS, H$_2$C$^{33}$S, H$_2$C$^{34}$S,
H$_2$C$^{36}$S, D$_2$CS, H$_2^{13}$C$^{34}$S \\

HNCO &
DNCO, H$^{15}$NCO, HN$^{13}$CO, HNC$^{17}$O, HNC$^{18}$O,
DN$^{13}$CO, HN$^{15}$CO, H$^{15}$N$^{13}$CO \\

HNCS &
DNCS, H$^{15}$NCS, HN$^{13}$CS, HNC$^{33}$S, HNC$^{34}$S,
DN$^{13}$CS \\

t-HCOOH &
t-DCOOH, t-HCOOD, H$^{13}$COOH, HCO$^{18}$OH, HC$^{18}$OOH,
t-DCOOD, t-$^{13}$COOH, t-HC$^{18}$O$^{18}$OH \\

HCS$^+$ &
DCS$^+$, H$^{13}$CS$^+$, HC$^{33}$S$^+$, HC$^{34}$S$^+$,
D$^{13}$CS$^+$ \\

\midrule
\multicolumn{2}{@{}l}{\textit{Carbon chains and complex organic molecules}}\\
\midrule


C$_2$H &
$^{13}$CCH, C$^{13}$CH, C$_2$D \\

l-C$_3$H &
l-C$_3$D, $^{13}$CC$_2$H, C$^{13}$CH, CC$^{13}$H \\

c-C$_3$H &
c-C$_3$D, c-$^{13}$CC$_2$H \\

C$_4$H &
C$_4$D, $^{13}$C-substituted C$_4$H (all four C positions) \\

C$_5$H &
C$_5$D, $^{13}$C-substituted C$_5$H \\

C$_3$N &
C$_3^{15}$N, $^{13}$CCCN, C$^{13}$CCN, CC$^{13}$CN \\

H$_2$C$_3$ &
H$_2$C$_2^{13}$C, H$_2^{13}$CC$_2$ \\

H$_2$C$_4$ &
H$_2$C$_3^{13}$C, H$_2$C$_2^{13}$CC \\

HC$_3$N &
HC$_3^{15}$N;
$^{13}$C-substituted HC$_3$N (three C positions);
$^{13}$C$_2$-substituted HC$_3$N (three combinations);
H$^{13}$C$_3$N;
$^{13}$C+$^{15}$N substituted HC$_3$N;
D-substituted HC$_3$N and corresponding $^{13}$C+$^{15}$N variants \\

HC$_5$N &
DC$_5$N, HC$_5^{15}$N;
$^{13}$C-substituted HC$_5$N (five C positions);
D+$^{13}$C substituted HC$_5$N;
DC$_5^{15}$N \\

HC$_7$N &
DC$_7$N, $^{13}$C-substituted HC$_7$N (seven C positions) \\

HC$_9$N &
DC$_9$N, $^{13}$C-substituted HC$_9$N (nine C positions) \\

c-C$_3$H$_2$ &
c-C$_3$HD, c-C$_3$D$_2$,
c-$^{13}$CC$_2$H$_2$, c-CC$^{13}$CH$_2$,
c-$^{13}$CC$_2$HD, c-CC$^{13}$CHD \\

CH$_3$CCH &
CH$_3$CCD, CH$_2$DCCH,
$^{13}$CH$_3$CCH, CH$_3^{13}$CCH, CH$_3$CC$^{13}$H \\

CH$_3$OH &
CH$_3$OD, CH$_2$DOH, CH$_3^{18}$OH,
$^{13}$CH$_3$OD, $^{13}$CH$_3^{18}$OH,
CHD$_2$OH, CD$_3$OH, CD$_3$OD,
$^{13}$CD$_3$OH, $^{13}$CD$_3$OD \\

CH$_3$CN &
CH$_2$DCN, $^{13}$CH$_3$CN, CH$_3^{13}$CN, CH$_3$C$^{15}$N,
CHD$_2$CN, CD$_3$CN,
$^{13}$CH$_3^{13}$CN, $^{13}$CH$_3$C$^{15}$N,
CH$_3^{13}$C$^{15}$N, CD$_3^{13}$CN \\

CH$_3$SH &
CH$_2$DSH, CH$_3$SD, CH$_2$DSH,
$^{13}$CH$_3$SH, CH$_3^{34}$SH \\

CH$_3$NH$_2$ &
$^{13}$CH$_3$NH$_2$ \\

CH$_3$CHO &
CH$_2$DCHO, CH$_3$CDO, $^{13}$CH$_3$CHO,
CH$_3^{13}$CHO, CH$_3$CH$^{18}$O, CD$_3$CHO,
CH$_3^{13}$CDO, $^{13}$CH$_3^{13}$CHO,
CD$_3$CDO, CHD$_2$CHO \\

NH$_2$CHO &
NHDCHO, NH$_2$CDO, $^{13}$NH$_2$CHO,
NH$_2^{13}$CHO, $^{15}$NH$_2$CHO, NH$_2$CH$^{18}$O \\

HOCH$_2$CN &
DOCH$_2$CN, HO$^{13}$CH$_2$CN, HOCH$_2^{13}$CN \\

CH$_3$OCH$_3$ &
CH$_3$OCH$_2$D, $^{13}$CH$_3$OCH$_3$,
CH$_3{}^{18}$OCH$_3$, CH$_3$OCHD$_2$ \\

HCOOCH$_3$ &
DCOOCH$_3$, HCOOCH$_2$D, H$^{13}$COOCH$_3$,
HCOO$^{13}$CH$_3$, HCO$^{18}$OCH$_3$,
HC$^{18}$OOCH$_3$, HCOOCHD$_2$ \\

CH$_3$CH$_2$OH &
a-CH$_3$CHDOH, s-CH$_3$CHDOH,
CH$_2$DCH$_2$OH, $^{13}$CH$_3$CH$_2$OH,
CH$_3^{13}$CH$_2$OH, CH$_3$CH$_2^{18}$OH,
CH$_3$CH$_2$OD \\

CH$_3$COCH$_3$ &
CH$_3$COCH$_2$D, $^{13}$CH$_3$COCH$_3$,
CH$_3{}^{13}$COCH$_3$ \\
\bottomrule
\end{tabularx}

\vspace{1pt}

\begin{minipage}{0.98\linewidth}
\tiny
\textit{Note.} Molecules in boldface are species for which the parent molecule
is symmetric, with no permanent dipole moment, and therefore not detectable at
millimeter wavelengths, whereas the listed isotopologues are detectable. O$_{2}$ is an exception since in principle is detectable at (sub)mm wavelengths through magnetic-dipole rotational transitions. 
Adapted from CDMS (\citealt{endres2016}), JPL (\citealt{pickett1998}), and
the Lille Spectroscopic Database (\citealt{motiyenko2025}).
\end{minipage}

\end{table}
\end{landscape}

\newpage{}
\noindent\textbf{Affiliations}

{
\noindent$^1$ Centro de Astrobiología (CAB), CSIC-INTA, Carretera de Ajalvir km 4, Torrejón de Ardoz, 28850 Madrid, Spain \Letter~\url{lcolzi@cab.inta-csic.es} ;\\
$^2$ Max-Planck-Institut f\"ur Extraterrestrische Physik (MPE), Giessenbachstr. 1, 85748 Garching, Germany; \\
$^3$ Department of Astronomy, University of Virginia, Charlottesville, VA22902, USA; \\
$^4$ INAF, Osservatorio Astrofisico di Arcetri, Largo E. Fermi 5, I-50125 Firenze, Italy; \\
$^5$ LUX, Observatoire de Paris, PSL Research University, CNRS, Sorbonne Université, 75014 Paris, France; \\
$^6$ UK Astronomy Technology Centre, Royal Observatory Edinburgh, Blackford Hill, Edinburgh EH9 3HJ, UK; \\
$^7$ Niels Bohr Institute, University of Copenhagen, Jagtvej 155A, 2200 Copenhagen N, Denmark; \\
$^{8}$ Planet and Space Chemistry Branch, NASA Goddard Space Flight Center, 8800 Greenbelt Road, Greenbelt, MD 20771, USA; \\
$^{9}$ Department of Physics, Catholic University of America, Washington, DC 20064, USA; \\
$^{10}$ Université de Toulouse, CNES, CNRS, IRAP, Toulouse, France; \\
$^{11}$ Laboratoire d’astrophysique de Bordeaux, Univ. Bordeaux, CNRS, B18N, 33615 Pessac, France; \\
$^{12}$European Southern Observatory, Karl-Schwarzschild-Straße 2, D-85478 Garching bei München, Germany; \\
$^{13}$ Department of Chemistry, Biochemistry and Pharmaceutical Sciences (DCBP), Universität Bern, Freiestrasse 3, 3012 Bern, Switzerland; \\
$^{14}$ Cardiff Hub for Astrophysics Research and Technology (CHART), PACES, Department of Physics and Astronomy, Cardiff University, The Parade, CF24 3AA Cardiff, UK; \\
$^{15}$ Chinese Academy of Sciences South America Center for Astronomy, NAO, CAS, Beijing 100101, China; \\
$^{16}$ Instituto de Astronom\'ia, Universidad Cat\'olica del Norte, Av. Angamos 0610, Antofagasta, Chile; \\
$^{17}$ Shanghai Astronomical Observatory, Chinese Academy of Sciences, 80 Nandan Road, Shanghai 200030, P. R. China; \\
$^{18}$ Univ. Grenoble Alpes, CNRS, IPAG, 38000 Grenoble, France; \\
$^{19}$ European Southern Observatory, Alonso de Córdova, 3107, Vitacura, Santiago 763-0355, Chile; \\
$^{20}$ Joint ALMA Observatory, Alonso de Córdova, 3107, Vitacura, Santiago 763-0355, Chile; \\
$^{21}$ INAF- Istituto di Radioastronomia, Via P. Gobetti 101, 40129, Bologna, Italy; \\
$^{22}$ Star and Planet Formation Laboratory, PRI, RIKEN, 2-1 Hirosawa, Wako, Saitama 351-0198, Japan;\\
$^{23}$ Institut de Ciències de l’Espai (ICE), CSIC, Campus UAB, Carrer de Can Magrans s/n, Bellaterra, E-08193 Barcelona, Spain; \\
$^{24}$~Departamento de Fisica aplicada III, ETSI Universidad de Sevilla, 41092 Sevilla, Spain; \\
$^{25}$ ESA/AURA, Space Telescope Science Institute, Baltimore, USA; \\
$^{26}$ Observatorio Astronomico Nacional (IGN), Spain; \\
$^{27}$ Instituto de Física Fundamental (CSIC), Calle Serrano 121-123, E-28006 Madrid, Spain; \\
$^{28}$ IAPS-INAF, Via Fosso del Cavaliere, 100, I-00133 Rome, Italy.\\

}

\end{document}